\documentclass[pra,paperpaper,twocolumn,showpacs,superscriptaddress,floatfix]{revtex4}

\usepackage{graphicx}
\usepackage{epstopdf}
\usepackage{dcolumn}
\usepackage{bm}

\usepackage{color}

\begin{document}

\title{Monopole excitations of a harmonically trapped one-dimensional Bose gas from the ideal gas to the Tonks-Girardeau regime}
\author{S. Choi}
\affiliation{Department of Physics, University of Massachusetts, Boston, MA 02125, USA}
\author{V. Dunjko}
\affiliation{Department of Physics, University of Massachusetts, Boston, MA 02125, USA}
\author{Z. D. Zhang}
\affiliation{Department of Physics and Astronomy, SUNY, Stony Brook, NY 11794  USA}
\affiliation{Department of Physics, University of Massachusetts, Boston, MA 02125, USA}
\author{M. Olshanii} 
\affiliation{Department of Physics, University of Massachusetts, Boston, MA 02125, USA}

\begin{abstract}
Using a time-dependent modified nonlinear Schr\"{o}dinger equation 
(m-NLSE) --- where the conventional chemical potential proportional to the density is replaced by 
the one inferred from Lieb-Liniger's exact solution --- we study frequencies of the collective 
monopole excitations of a one-dimensional (1D) Bose gas.    
We find that our method accurately reproduces the results of a recent experimental study 
[E. Haller {\it et al}., Science \textbf{325}, 1224 (2009)] 
in the full spectrum of interaction regimes from the ideal gas, through the mean-field regime, through the 
mean-field Thomas-Fermi regime, all the way to  
the Tonks-Giradeau gas. While the former two are accessible 
by the standard time-dependent NLSE and inaccessible by the time-dependent local density approximation (LDA), 
the situation {\it reverses} in the latter case. However, 
the m-NLSE is shown to treat all these regimes 
within a single numerical method. 
\end{abstract}

\pacs{67.85.-d, 02.60.Cb}

\maketitle

The study of  excitations of a material  allows us to understand its underlying nature and forms the basis of 
various spectroscopic methods.  In particular, collective excitations of ultracold atoms provide a way to infer  
their  character, including the nature of their interatomic interactions.  The goal of this paper 
is to simulate the monopole oscillations for a Bose gas in a one-dimensional (1D) harmonic oscillator 
(HO)  potential for all range of interaction strengths to  demonstrate the continuous transition from the 
bosonic ideal gas and mean-field regimes at weak interaction to the
fermionic strongly correlated limit at large interaction strength.  A number of experiments on the excitations of 1D bose gas exist\cite{Moritz,Nagerl,Fang}, and yet a unified theoretical description over all interaction regimes has not been available.

Previous work by various authors in describing 1D Bose gas all have involved either some form of Nonlinear Schr\"odinger Equation or the equivalent hydrodynamic formulation which are intimately connected via the Madelung transformation as we discuss below. It is found that these have generally been fragmented in terms of the applicable range of interactions. 
For weakly interacting 1D Bose gas,  the 1D Gross-Pitaevskii Equation (GPE) applies, 
with the coupling constant given by  $g_{1D} = -\frac{2\hbar^2}{ma_{1D}}$, 
where $a_{1D} = (-a_{\bot}^2/2a) [1 -  C(a/a_{\bot})]$ is the 1D scattering length (negative for repulsive 
3D interactions and weak transverse confinement, i.e.\ for $0 < a < C a_{\bot}$),  $a_{\bot} = \sqrt{2\hbar/m\omega_{\bot}}$ is the size of the transverse 
ground state wave function, $\omega_{\bot}$ is the frequency of the transverse confinement, $C \equiv -\zeta(1/2)= 1.4603\ldots$, and $\zeta(z)$ is Riemann's 
zeta-function \cite{olshanii}. The above expression takes in to account a possibility of a virtual excitation of higher modes in the confining direction at the moment 
of a two-body collision. 
For the case when the mean-field potential becomes comparable to the transverse excitation quanta, Salasnich {\it et al.}\cite{Salasnich2002} 
derive a cubic nonlinear Schr\"{o}dinger Equation (NLSE) that includes the effect of ``spilling'' of the 
transverse component of the 3D GPE wavefunction to higher transverse modes. 
It is found that the resulting equation can describe both the weakly interacting mean field and strongly interacting high density Thomas-Fermi (TF) 1D bosons, but not the strongly interacting low atom density Tonks-Girardeau (TG)  regime. For the TG regime where interacting impenetrable bosons in 1D behave as noninteracting fermions, a higher nonlinearity than the usual cubic one of GPE is required, and  Kolomeisky {\it et al.}\cite{Kolomeisky2000} derive a NLSE that includes quintic nonlinearity. 
It is notable that Minguzzi {\it et al.}\cite{Minguzzi2001} derived from the quintic NLSE of Kolomeisky {\it et al.}\cite{Kolomeisky2000}, Landau's hydrodynamic equation that matches the well-known result for noninteracting Fermi gas in the hydrodynamic regime\cite{March1972}.

In general, the hydrodynamic equations were found to work beyond mean field\cite{PitString,Merloti}, including the TG regime\cite{MenString}. The hydrodynamic equations are:
 \begin{equation}
  \frac{\partial}{\partial t} n + {\bf \nabla} ({\bf v} n ) = 0 
  \label{continuity}
  \end{equation}
 \begin{equation}  
  m \frac{\partial}{\partial t} {\bf v} + {\bf \nabla} \left ( \mu + V_{ext}({\bf r}) + \frac{1}{2} m{\bf v}^2  \right ) =  0
  \label{Euler}
  \end{equation}
 where $n$ is the density of gas, ${\bf v}$ is the velocity field, $\mu = \mu_{l}[n({\bf r},t)]$ is the 
local density-dependent chemical potential,  $\mu_{l}$ is the chemical potential calculated for a {\it uniform} gas at density $n({\bf r},t)$, and $V_{ext}({\bf r})$ is the external confining potential.  
The hydrodynamic equations involve the local density approximation (LDA) that corresponds to the case of zero temperature, large $N$ limit, and ``macroscopic'' dynamics where length scales are much larger than both the interparticle distance and the healing length. 
These conditions, although they superficially appear strict, are found not too difficult to meet in practice. 
 We point out here that the connection between the hydrodynamic equations and NLSE is well-established\cite{Stringari1,Stringari2}. In going from NLSE to the equivalent hydrodynamic formulation, an additional quantum pressure term, $ {\bf \nabla} (-\frac{\hbar^2}{2m \sqrt{n}} {\bf \nabla}^2 \sqrt{n})$, is necessarily introduced in Eq. (\ref{Euler}).  For sufficiently smooth density distributions, such quantum pressure term is inconsequential on length scales that are much larger than the characteristic microscopic length scales of the problem (such as the healing length, or mean interparticle separation), and the NLSE and the hydrodynamic equations are practically and computationally equivalent in such cases. 
   
Some of us\cite{DunLorOls} used the hydrodynamic equations to study the stationary state of 1D bose gas in HO potential  from the mean field to the TG regimes via the calculation of required chemical potential encompassing all regimes of quantum correlations. In order to find the chemical potential the 
Lieb-Liniger Hamiltonian with zero range 1D repulsive potentials,
$
\hat{H} =  -\frac{\hbar^2}{2m} \sum_{i = 1}^{N} \frac{\partial^2}{\partial {z_i}^2} +   g_{1D} \sum_{i < j}^{N} \delta(z_i - z_j)
$, was used to calculate the chemical potential for this system using  $\mu_{l} = \partial [n \epsilon(n)]/ \partial n $ where the energy per particle $\epsilon(n) = \hbar^2 n^2 e(\gamma(n))/2m$ comes from  solving  the Lieb-Liniger system of equations  that arise from applying the Bethe Ansatz\cite{liebliniger}. Here, $\gamma(n)$ is the dimensionless Lieb-Liniger interaction parameter proportional to the interaction strength $g_{1D}$:  $\gamma(n) = 2/n|a_{1D}|$.   
 
\"{O}berg and Santos\cite{Ohberg2002} and Pedri {\it et al.}\cite{Pedri2003}  extended Ref. \cite{DunLorOls} to study the free expansion of 1D bose gas when the harmonic trap is released, by converting the hydrodynamic equations to a NLSE, via the inverse Madelung transform\cite{Stringari1,Stringari2}. 
The work of Ref. \cite{Ohberg2002} was  limited to narrow interaction regime between TF and TG, as characterized by the interaction parameter $\eta = n^{0}_{TF}|a_{1D}|  \approx 1$ where $n^{0}_{TF} = [(9/64)N^2|a_{1D}|/a_{z}^4]^{1/3}$ is TF density with $a_z = \sqrt{\hbar/m \omega_z}$. In addition, owing to the specific form of the ansatz for their wave function, it is not {\it a priori} obvious whether the work may be extended beyond free-expansion that they studied.  Indeed, more sophisticated experiment beyond free expansion, in particular  that of  measuring the monopole oscillation frequency from the mean field to the TG regime has been performed\cite{Nagerl}, where
the interaction of  8 to 25 ultracold Cs atoms trapped in effectively 1D harmonic trap was tuned via Feshbach resonance while measuring the change in the ratio of the oscillation frequencies of the collective compression  ($\omega_{C}$) and dipole ($\omega_{D}$) modes, $R = (\omega_{C}/\omega_{D})^2$, the change of which provides the diagnostics for the crossover between different regimes. 

So far the experimental regime between TF and TG that shows the crossover of the oscillation frequency has been described by the hydrodynamic equations  combined with sum rules\cite{MenString} which  calculate the {\it upper limit} to the excitation based on the static wave functions. The sum rules method, however, cannot simulate directly the {\it dynamics} of the 1D bose gas. Very recently, it was shown that a Hartree approach allows for an accurate description of Gaussian Bose-Einstein condensate (BEC) to TF regimes and shown to join smoothly to the crossover from TF to TG described using LDA for the cases involving more than $25$ particles\cite{Gudyma}. It is further shown that  {\it ab initio} diffusion Monte Carlo calculations provide a complete all-regimes data for the cases involving less than 25 atoms.

These limitations are expected to be overcome in time with, for instance, improved numerical methods and more powerful computer. In this Letter, we show how we overcame these limitations using the modified NLSE (m-NLSE) by the inverse Madelung transform, starting with the hydrodynamic equations and adding in the quantum pressure term as we discussed above for mathematical consistency. To obtain m-NLSE, the standard NLSE nonlinear term, $g_{1D}|\Psi({\bf r}, t)|^2 \Psi({\bf r}, t)$ is replaced by $\mu(n\!=\!|\Psi({\bf r}, t)|^2)\Psi({\bf r}, t)$.  
Working with a wavefunction $\psi({\bf r}, t) = N^{-1/2} \Psi({\bf r}, t)$ normalized to unity, $\int_{-\infty}^{+\infty} |\psi|^2 \,dz = 1$,  and, accordingly,  with a probability distribution $\tilde{n} = n/N = |\psi|^2 $ and the external potential given by a one-dimensional harmonic potential $V_{ext}({\bf r}) = \frac{1}{2} m \omega_{z}^2 z^2$ The modified NLSE (m-NLSE) in the harmonic oscillator system of units: $\hbar=m=\omega=1$ is
\begin{equation}
i \frac{\partial \psi(z)}{\partial t} = \left [ -\frac{1}{2} \frac{\partial^2}{\partial z^2} +
     \frac{1}{2} z^2  + \mu[\tilde{n}(z, t)] \right ] \psi(z)  
\label{GPE_general}
\end{equation}
where the chemical potential is given by:
$
 \mu[\tilde{n}(z, t)] = \frac{N^2}{2} \tilde{n}^2  \left (3 + \tilde{n} \frac{\partial  }{ \partial \tilde{n}} \right ) e(\gamma(
 n\!=\! N\tilde{n}))   \label{GPE_LLchempot}
$ calculated numerically from the Lieb-Liniger system of equations\cite{liebliniger} at each spatio-temporal step as $\tilde{n}(z,t)$ evolves. We note that these equations have indeed been obtained earlier\cite{Ohberg2002}. In terms of the validity of the m-NLSE, from the mathematical assumptions made, 
the equation should be valid for all situations where both the LDA and the Lieb-Liniger theory hold. It is noted additionally that throughout the whole range of the interaction strength, the quantum pressure term is either exact (as in the ideal gas and in the GPE regimes, the regime of no interactions, and the regime of weak but non-negligible interactions, the latter thanks to the presence of BEC) or negligible (as in the Tonks-Girardeau regime, the regime of strong interactions).   

We note that  in the mean field limit of $n a_{1D} \gg 1$, $\gamma \rightarrow 0$  leading to the well known TF energy functional; there, the chemical potential is given by $\mu(n) \approx g_{1D} n = \gamma \hbar^2n^2 /m$. In the TG limit of  $n a_{1D} \ll 1$ where $\gamma \rightarrow \infty$,  the chemical potential is $\mu(n) \approx  \pi^2 \hbar^2 n^2/2m$. In between the TF and TG limits the chemical potential has to be worked out numerically. 
For comparison with experiments, it is convenient to define the effective $\gamma$ instead of  $\gamma(n)$ via the maximum steady state density of the atomic cloud at $z = 0$ in the TG limit and the actual $a_{1D}$:
\begin{eqnarray*}
\gamma_{eff.} = \frac{2}{n_{TG}(0) |a_{1D}|}  = \frac{\pi}{\alpha}
\end{eqnarray*}
where $n_{TG}(0) = \sqrt{2N m \omega_z/\hbar}/\pi$ is the analytical TG density in the center of the trap\cite{DunLorOls}, and for convenience we defined dimensionless parameter $\alpha =  \sqrt{ N m \omega_z /2\hbar }  |a_{1D}| $ that parametrizes the regimes of interaction strength. This naturally introduces a set of $\gamma$ independent of the density profile, and we shall use $\gamma_{eff.}$ as our parameter in our simulation. 



The simulation was done by first finding the ground state solution for various values of $\alpha$ starting from the ground state for a HO i.e. zero effective chemical potential $\mu[\tilde{n}(z, t)]$ in Eq. (\ref{GPE_general}) with $N =0$ and adiabatically increasing $N$ up to the desired number of atoms. The idea follows from the well-known quantum mechanical theorem on adiabatic following\cite{AdiabaticFollowing}, which is the limiting case of Landau-Zenner transition with zero transition probability such that a system remains in the state that evolves from the initial state in the limit of infinitely slow evolution of the time-dependent Hamiltonian. This turned out to be a crucial numerical step since the strong nonlinearity makes direct numerical solution to the ground state i.e. not via some kind of variational ansatz difficult. In this case, the imaginary time evolution or the damping method\cite{Choi} to obtain the ground state was also found to run into convergence problems, possibly due to highly nonlinear energy landscape. The fact that the adiabatic following method works well indicates that m-NLSE may be applied to simulations involving general trapping potential other than harmonic, as long as the corresponding non-interacting ground state is known.

Once the steady state solutions are found, the monopole excitation can then be simulated in many different ways, including an addition by hand of the exact Bogoliubov excitation modes or sinusoidal driving of the confining potential.  In this paper, we directly excite monopole oscillations by suddenly quenching the confining potential, from $V(z) = \frac{1}{2} m \omega_{z}^2 z^2$ to  $V(z) = 1.25 \times \frac{1}{2} m \omega_{z}^2 z^2$ for some short time ($\tau = 0.25 \pi/\omega_{z}$) then back in the original trap frequency. The simulation was then run until $\tau = 400 \pi/\omega_{z}$ while measuring the time-dependent width (variance)  of the wave function 
$\langle \Delta z^2 \rangle = \int z^2 |\psi(z,t)|^2 d z  - \left [ \int z |\psi(z,t)|^2 dz \right ]^2$.
It was found that except for a short transient, the width $\langle \Delta z^2 \rangle$ follows a sinusoidal variation over time, owing to the harmonic confining potential. From this sinusoidally varying time-dependent width, the Fourier frequency components were obtained numerically.  

\begin{figure}[t] \begin{center}
{\includegraphics[height=6cm]{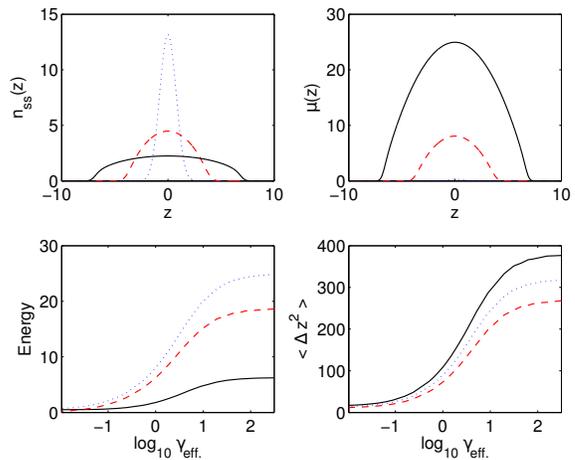}}
\caption{Top left panel: Plot of the steady state density $n_{ss}(z) = |\psi(z)|^2$ vs. $z$ with atom number $N = 25$ for  $\gamma_{eff.} \approx 0.01$ (dotted line), 1  (dashed line), and $\gamma_{eff.} \rightarrow \infty$ (solid line)  Top right panel:  the corresponding chemical potential $\mu(z)$.  It is noted that  $\mu(z)$ for  $\gamma_{eff.} \approx 0.01$ (dotted line) is vanishingly small.  Bottom left panel: Plot of the harmonic oscillator energy $E_{HO}$ (solid line), interaction energy $E_I$ (dashed line) and the total energy $E_K + E_I$ (dotted line) as defined in the text as a function of $\log_{10} \gamma_{eff.}$  Bottom right panel:  the initial width of the wave function $\langle \Delta z^2 \rangle$  (dotted line),   minimum $\langle \Delta z^2 \rangle$  (dashed line),  and maximum $\langle \Delta z^2 \rangle$   (solid line) attained during the monopole oscillation as a function of $\log_{10} \gamma_{eff.}$.  
The harmonic oscillator units are used throughout.}   \label{ProfileEnergy}
\end{center}
\vspace{-0.75cm}
\end{figure}

We plot in Fig. \ref{ProfileEnergy}   the steady state density $n_{ss}(z)$  with atom number $N = 25$ for  $\gamma_{eff.} \approx 0.01$, 1, and  $\gamma_{eff.} \rightarrow \infty$ and the corresponding chemical potential $\mu(z)$.   
The position-dependent chemical potential gives an idea of the effective potential experienced by the wave function due to interatomic interaction. This function was found to almost vanish for $\gamma_{eff.} < 1$, leaving an effectively  interaction-free system of atoms. We also plot the steady state harmonic oscillator energy $E_{HO} = \int   \psi^{*}(z)  \left [ -\frac{1}{2} \frac{\partial^2}{\partial z^2} + \frac{1}{2} z^2  \right ] \psi(z)  d z$, interaction energy $E_I =  \int  \psi^{*}(z) \mu(z) \psi(z)  d z$,  and the total energy $E_K + E_I$ as a function of $\log_{10} \gamma_{eff.}$.  Additionally, we plot  as a function of $\log_{10} \gamma_{eff.}$  the initial, maximum and minimum width of the wave function $\langle \Delta z^2 \rangle$  attained during the monopole oscillation. It is not surprising that the initial width of the wave function  follows the trend of the total energy of the system as a function of $\gamma_{eff.}$ since the increasing repulsion between the atoms makes the wave function profile wider.  The amplitude of oscillation is seen to also grow as a function of $\gamma_{eff.}$;  however taking into account the change in the initial width itself the oscillation amplitude  remains constant at approximately 20\% of the initial width regardless of $\gamma_{eff.}$.  

\begin{figure}[t] 
\vspace{0.75cm}
\begin{center}
{\includegraphics[height=6cm]{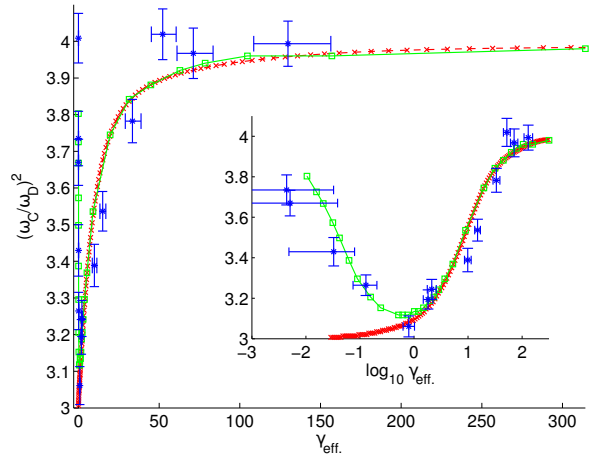}}
\caption{(Color online) Plot  of the monopole oscillation frequency squared $(\omega_{C}/\omega_{D})^2$  as a function of $\gamma_{eff.} = 2 \pi/\sqrt{2N m \omega_z/\hbar} |a_{1D}|$.  Inset: same figure as a function of  $\log_{10} \gamma_{eff.}$ to magnify the region near $\gamma_{eff.} \approx 0$---the region that lies 
{\it outside} of the range of validity of hydrodynamic equations. Stars with error bars: the experimental data from N\"{a}gerl group\cite{Nagerl}. Green squares: numerical simulation, Red crosses:  sum rules result \cite{MenString}.
It must be mentioned that in the region of $\gamma_{eff.} \ll 1$, the result \cite{MenString} is not expected to 
match the experiment since it has been built on the hydrodynamic equations {\it a priori}. The difference may be understood from the fact that the effect of the quantum pressure term added in deriving the m-NLSE becomes more significant in the regime of weak interactions.
}  \label{Experimental}
\end{center}
\vspace{-0.75cm}
\end{figure}

Our numerical simulation parameters are within the range of experimental parameters:  we cover the same range of $\gamma_{eff.}$  as in the experiment and we use $N = 25$.   In Fig. \ref{Experimental} we plot the experimentally measured frequencies $(\omega_{C}/\omega_{D})^2$ with error bars as a function of $\gamma_{eff.}$  and superpose the results from our simulation as well as the  prediction.  The near-ideal gas region corresponds to the frequency interval from $(\omega_{C}/\omega_{D})^2  \approx  4$ to $3$. The point $(\omega_{C}/\omega_{D})^2  \approx  3$ ($\gamma  \approx 1$) is the mean-field TF point. For higher $\gamma_{eff.}$, the system slowly approaches a TG plateau of $(\omega_{C}/\omega_{D})^2  \approx  2$. The sum rule formula of Ref.~\cite{MenString},
which was built using hydrodynamic equations, works well in both mean-field TF and in the TG regimes, but naturally fails for the 
near-ideal gas. Our approach however captures it. The difference may be understood from the fact that the effect of the quantum pressure term added in deriving the m-NLSE becomes more significant in the regime of weak interactions. 

We note that the sum rules do not necessarily require LDA; the only calculation available for meaningful comparison, Ref. \cite{MenString}, just happens to be built on LDA as they were more interested in the $\gamma \rightarrow \infty$ region. 
Granted, our method is more than a naive interpolation between the standard NLSE at weak interactions and the LDA at the strong ones, since there exists a parameter region---the mean-field TF regime---of {\it overlapping ranges of validity} of the above methods.
Note that while at the ideal gas point and in the subsequent mean-field regime, the m-NLSE correctly describes the density evolution at all length scales, in the strongly correlated regime, the m-NLSE must be regarded merely as a simulator for the time-dependent LDA equations which is convenient since one does not have to simultaneously track the velocity field and the density -- it suffices to track a single wave function, and any features of a healing length size or smaller must be treated as artifacts of the computational method. This question is discussed in Ref.\cite{Girardeau2000}: it is shown in particular that the interference fringes produced by the m-NLSE in the TG regime\cite{Kolomeisky2000} have nothing to do with reality.

In conclusion, we found that using a single m-NLSE one can consistently simulate the 1D Bose gas in the full spectrum of interaction regimes. Besides being numerically tractable ({e.g.} the sharp edges of the atomic clouds are automatically regularized), the m-NLSE offers the following benefit: At very low densities, where the size of the cloud become comparable to the size of the one-body quantum ground state of the trap, the time-dependent LDA fails while the standard NLSE is naturally valid there; but this is exactly what the m-NLSE converts to and so m-NLSE is able to capture the system's behavior at very low values of the interaction strength. This allows for a formal numerical unification of the standard time-dependent NLSE valid in the ideal gas limit and in the neighboring mean-field regime (both before the validity of the TF approximation and in the TF regime) and the time-dependent LDA valid for the mean-field TF regime, TG regime, and in between. 

The m-NLSE bridges the gap between existing previous work Refs.  \cite{olshanii}, \cite{Salasnich2002}, \cite{Kolomeisky2000}, \cite{MenString}, \cite{DunLorOls}, and \cite{Ohberg2002} each of which has restricted range of applicability.   Also unlike Refs. \cite{MenString} and \cite{DunLorOls} the dynamics can be simulated directly. The numerically intensive diffusion Monte Carlo method reported very recently can simulate over all interaction regimes but is restricted to small number of atoms\cite{Gudyma}.  
Furthermore, the adiabatic following method for the ground state preparation implies potential for application of m-NLSE to a broad range of future research. On the other hand, we reiterate that there are certain obvious limitations of m-NLSE/hydrodynamic approach such as that discussed by Girardeau and Wright in Ref.\cite{Girardeau2000}, to do with phenomena at healing length scales.

 In general, one may safely apply the m-NLSE wherever the LDA holds;  many experimental situations involving large amplitude motion, such as problems in quantum transport should satisfy the LDA  and hence render the m-NLSE a fully valid theoretical model.  Although the range of validity of m-NLSE is clear from considering the underlying mathematical assumptions, it is also possible that, just as GPE with its theoretically narrow range of applicability (zero temperature, mean field) found wide applications, the m-NLSE may have a broader applicability than expected\cite{Zhang}. In this sense, more experiments are needed to be done and compared with our m-NLSE to establish the range of validity.
On a more fundamental level, since m-NLSE is a numerical tool for simulating the hydrodynamic equations, future research should involve careful examination of the validity of the hydrodynamic equations themselves, along the lines of Ref. \cite{Zaremba1988,Braaten1999}.

\end{document}